\documentclass[sigconf]{acmart}

\AtBeginDocument{%
  \providecommand\BibTeX{{%
    \normalfont B\kern-0.5em{\scshape i\kern-0.25em b}\kern-0.8em\TeX}}}

\copyrightyear{2022}
\acmYear{2022}
\setcopyright{acmcopyright}
\acmConference[ICSE '22]{44th International Conference on Software Engineering}{May 21--29, 2022}{Pittsburgh, PA, USA}
\acmBooktitle{44th International Conference on Software Engineering (ICSE '22), May 21--29, 2022, Pittsburgh, PA, USA}
\acmPrice{15.00}
\acmDOI{10.1145/3510003.3510116}
\acmISBN{978-1-4503-9221-1/22/05}

\usepackage{tcolorbox}
\usepackage{multirow}
\usepackage{graphicx}
\usepackage{subcaption}
\usepackage{url}
\usepackage[shortlabels]{enumitem}

\definecolor{green}{HTML}{05AA4F}
\definecolor{yellow}{HTML}{FFEA3D}
\definecolor{blue}{HTML}{102694}
\definecolor{magenta}{HTML}{FB3199}
\definecolor{brown}{HTML}{782106}
\definecolor{yellowgreen}{HTML}{92D072}
\definecolor{orange}{HTML}{FF7B37}
\definecolor{skyblue}{HTML}{00CEE1}
\definecolor{red}{HTML}{FF2C16}
\definecolor{apricot}{HTML}{FFB781}

\begin{document}

\title{GitHub Sponsors:\\Exploring a New Way to Contribute to Open Source}

\author{Naomichi Shimada}
\affiliation{%
  \country{Nara Institute of Science and Technology, Japan}
}
\email{shimada.naomichi.sm3@is.naist.jp}

\author{Tao Xiao}
\affiliation{%
  \country{Nara Institute of Science and Technology, Japan}
}
\email{tao.xiao.ts2@is.naist.jp}

\author{Hideaki Hata}
\affiliation{%
  \institution{Shinshu University}
  \country{Japan}
}
\email{hata@shinshu-u.ac.jp}

\author{Christoph Treude}
\affiliation{%
  \institution{University of Melbourne}
  \country{Australia}
}
\email{christoph.treude@unimelb.edu.au}

\author{Kenichi Matsumoto}
\affiliation{%
  \country{Nara Institute of Science and Technology, Japan}
}
\email{matumoto@is.naist.jp}


\renewcommand{\shortauthors}{Shimada et al.}

\newcommand{\RqOne}{\textbf{RQ1:} \emph{Who participates in GitHub Sponsors?}}

\newcommand{\RqTwo}{\textbf{RQ2:} \emph{What characteristics make developers more likely to receive sponsorship?}}

\newcommand{\RqThree}{\textbf{RQ3:} \emph{What are developers' perceived challenges and benefits related to sponsoring?}}

\graphicspath{{figs/}}

\begin{abstract}
GitHub Sponsors, launched in 2019, enables donations to individual open source software (OSS) developers. Financial support for OSS maintainers and developers is a major issue in terms of sustaining OSS projects, and the ability to donate to individuals is expected to support the sustainability of developers, projects, and community. In this work, we conducted a mixed-methods study of GitHub Sponsors, including quantitative and qualitative analyses, to understand the characteristics of developers who are likely to receive donations and what developers think about donations to individuals. We found that: (1) sponsored developers are more active than non-sponsored developers, (2) the possibility to receive donations is related to whether there is someone in their community who is donating, and (3) developers are sponsoring as a new way to contribute to OSS. Our findings are the first step towards data-informed guidance for using GitHub Sponsors, opening up avenues for future work on this new way of financially sustaining the OSS community.
\end{abstract}

\keywords{GitHub Sponsors, Open Source, Sponsorship}


\maketitle

\section{Introduction}

Open source software (OSS) projects are often affected by economic challenges, and soliciting donations is a common means of obtaining funds for OSS projects.
However, as pointed out by Staltz, 
many OSS maintainers and developers do not generate enough income to sustain their OSS projects~\cite{sbtpl}.

In May 2019, GitHub introduced GitHub Sponsors~\cite{GHSponsors}, a service which allows OSS developers to accept donations from other GitHub users. While most OSS donation services in the past have targeted projects, GitHub Sponsors is unique in that it allows users to donate to individual OSS developers. GitHub advertises the service as ``a new way to contribute to open source'', in particular pointing out the widespread dependence on OSS projects: ``\textit{Sponsor the open source software your team has built its business on. Fund the projects that make up your software supply chain to improve its performance, reliability, and stability.}'' Anyone who contributes to an OSS project and lives in one of the 36 regions in which GitHub Sponsors was available at the time of writing is eligible to become a sponsored developer. Developers can join a waitlist, and once approved, a `Sponsor' button will appear on their profile and they can define sponsorship tiers. As of now, GitHub matches contributions of up to \$5k during a developer’s first year in GitHub Sponsors.

But who participates in GitHub Sponsors, what makes developers more likely to receive sponsorship, and what challenges and benefits do GitHub Sponsors bring? To answer such questions, we conducted a mixed-methods study of those who signed up for GitHub Sponsors and those who contributed donations.

Existing work on donations in the context of OSS projects---before the launch of GitHub Sponsors---reported that the activity and popularity of a project on GitHub are positively correlated with the likelihood of receiving donations, and projects that received donations show a short-term increase in the number of commits and a reduction in the time to resolve issues~\cite{10.1145/3377811.3380410}.
Krishnamurthy et al.~studied donations in SourceForge~\cite{RePEc:eee:respol:v:38:y:2009:i:2:p:404-414}. They reported that the decision to donate is impacted by relational commitment with the open source software platform, donation to projects, and accepting donations from others. Furthermore, the length of association with the platform and relational commitment affects donation levels.
Nakasai et al.~studied the impact of donation badges on developers' responses to Eclipse bug reports~\cite{8501934}.
They reported that bug reports from users who have donation badges used in Eclipse, take less time to receive a reply from the developers than other bug reports. Compared to these studies which investigated donations to projects, in this work, we study donations to individual developers, enabled by GitHub Sponsors.

From our quantitative analysis, we find that sponsored developers are more active than those looking for sponsorship unsuccessfully, and that many of the sponsors are active developers themselves, effectively building clusters of sponsorship. Distinguishing sponsored developers from non-sponsored developers, the presence of sponsorship in their development community is an important factor. Our qualitative analysis also reveals benefits and challenges related to GitHub Sponsors. Developers typically do not have channels at their disposal to attract sponsors and communicate with those who might be interested in donating, but on the other hand, many donate to show appreciation, not expecting anything specific in return. 

\paragraph{Significance of research contribution} Given the sustainability concerns affecting many OSS projects that are depended on by millions, understanding how sponsorship can provide the needed support is a crucial step towards ensuring that we can continue to rely on OSS projects. GitHub Sponsor's focus on enabling sponsoring of individual developers instead of projects is unique and has not been studied before. Our work provides insights into the characteristics, benefits, and challenges of the early adopters of GitHub Sponsors.


\section{Research Questions}
In this work, we conduct an exploratory study of GitHub Sponsors, a new approach to support developers.
%
Although GitHub Sponsors allows GitHub users to sponsor organizations as well as individuals, this study only focuses on the sponsorship of individual developers. 
Also, this study only targets sponsorship by individual developers and does not cover sponsorship by organizations.
The main goal of the study is to gain insights into the use of GitHub Sponsors among individuals.
Based on this goal, we constructed three main research questions to guide our study.
We present these main questions and sub-questions along with their motivation.

\vspace{2mm}
\RqOne
\begin{description}
    \item[RQ1.1]: What are the characteristics of sponsored developers?
    \item[RQ1.2]: What are the characteristics of sponsors?
\end{description}
The motivation of our first main research question \textbf{RQ1} is to 
set the stage for the rest of the paper by providing demographic information of involved receivers (\textbf{RQ1.1}) and sponsors (\textbf{RQ1.2}). 

\vspace{2mm}
\RqTwo \\
For OSS developers who want to obtain sponsorship, we set this second research question (\textbf{RQ2}) to identify the characteristics of developers who are likely to obtain sponsorship.

\vspace{2mm}
\RqThree
\begin{description}
    \item[RQ3.1]: Why are developers looking for sponsors?
    \item[RQ3.2]: What is the impact of (not) getting sponsorship?
    \item[RQ3.3]: Why are developers sponsoring?
\end{description}
To deepen our investigation of the answers to \textbf{RQ2}, we then ask the corresponding `why' questions for \textbf{RQ3}, again from both sides of a potential donation: why are developers looking for sponsorship (\textbf{RQ3.1}) and why are developers sponsoring (\textbf{RQ3.3})? We assumed that the main challenge related to sponsorship is not receiving money, so we asked about this explicitly in \textbf{RQ3.2}.

\section{Research Methods}

In this section, we describe our methods for data collection and analysis.

\subsection{Data Collection}
\label{ssec:data}

This section describes the steps to identify GitHub users who are participating in GitHub Sponsors, i.e., registered developers who are asking for sponsors, and GitHub users who are donating to developers.

\textbf{Preparing repositories.}
Since there is no direct way to get the users who participate in GitHub Sponsors, we first examined repositories to get the list of contributing GitHub users.
In order to cover as many users involved in GitHub sponsorship as possible, we collected a large number of GitHub repositories, i.e., repositories that were created between 2008 and 2020 and had 10 or more stars when we accessed them. 
We will discuss the limitations of this threshold of 10 stars in Section~\ref{ssec:t2v}.
Access and cloning was done between January and March 2021, and 1,210,041 repositories were collected. We excluded 387 repositories with no commits, resulting in 1,209,654 repositories.

\textbf{Obtaining contributors.}
The GitHub REST API was used to retrieve the contributors from each of the repositories collected; due to API limitations, the top 500 contributors from each repository were retrieved.
When we made requests via API between June and July 2021, there were repositories for which we could not get contributors because the repositories had been deleted or set to private, for example.
We obtained 1,695,015 contributors from 1,168,856 repositories.
The limitations of the threshold of the top 500 contributors are also discussed in Section~\ref{ssec:t2v}.

\textbf{Identifying GitHub Sponsors participants.}
In July 2021, We used the GitHub GraphQL API to identify developers who can be sponsored via GitHub Sponsors, i.e., developers who are seeking sponsorship. From the 1,695,015 contributors, we identified 9,366 such developers.
After analyzing each developer's user profile page for GitHub Sponsors and removing developers who had deleted their profile, we identified 3,697 developers who had obtained sponsors (\textbf{sponsored developers}) and 5,666 developers who had not yet obtained sponsors (\textbf{non-sponsored developers}).
%
From the analysis of a user profile page, we can get the list of sponsoring GitHub users including anonymous sponsors. Organization accounts were removed, and we identified a total of 17,458 non-anonymous individual sponsors. This includes sponsored and non-supported developers.




\textbf{Identifying primary programming languages.}
Since we have the list of contributors for all repositories we analyzed, we can aggregate information about the repositories contributed to by each developer. The primary language of the repositories can be retrieved using the GitHub GraphQL API, so the most common primary language of the repositories to which each developer contributed is taken as the primary language of that developer. 
Note that this is a rough estimate because we have not analyzed whether the developer commits in that language or not. The primary languages of developers identified in this way can be interpreted as the programming languages of the ecosystems to which the developers mainly contributed.

\subsection{Survey}
\label{ssec:survey}

\begin{table}
\caption{Survey participant demographics}
\label{tab:participant}
\small
\begin{tabular}{lrrrrrr}
\toprule
& \multicolumn{2}{c}{\textbf{sponsored}} & \multicolumn{2}{c}{\textbf{non-sponsored}} & \multicolumn{2}{c}{\textbf{sponsoring}} \\
\midrule
years coding \\
\cmidrule{1-1}
less than 5 & 7 & (4\%) & 4 & (3\%) & 4 & (6\%) \\
5-9 & 44 & (25\%) & 49 & (38\%) & 17 & (26\%) \\
10-19 & 73 & (41\%) & 47 & (37\%) & 25 & (38\%) \\
20-29 & 33 & (19\%) & 21 & (16\%) & 10 & (15\%) \\
30 or more & 20 & (11\%) & 8 & (6\%) & 10 & (15\%) \\
\midrule
years on GitHub \\
\cmidrule{1-1}
less than 5 & 23 & (13\%) & 18 & (14\%) & 15 & (23\%) \\
5-9 & 98 & (55\%) & 87 & (67\%) & 37 & (56\%) \\
10 or more & 56 & (32\%) & 24 & (19\%) & 14 & (21\%) \\
\midrule
gender \\
\cmidrule{1-1}
male & 157 & (89\%) & 111 & (86\%) & 60 & (91\%) \\
female & 6 & (3\%) & 8 & (6\%) & 3 & (5\%) \\
\multicolumn{6}{l}{non-binary, genderqueer, or gender non-conforming} \\
& 8 & (5\%) & 5 & (4\%) & 2 & (3\%) \\
prefer not to say & 6 & (3\%) & 5 & (4\%) & 1 & (2\%) \\
\midrule
OSS contributions \\
\cmidrule{1-1}
full time & 30 & (17\%) & 15 & (12\%) & 5 & (8\%) \\
part time & 129 & (73\%) & 96 & (74\%) & 44 & (67\%) \\
other & 18 & (10\%) & 18 & (14\%) & 17 & (25\%) \\
\midrule
\textbf{sum} & \textbf{177} & \textbf{(100\%)} & \textbf{129} & \textbf{(100\%)} & \textbf{66} & \textbf{(100\%)} \\
\bottomrule
\end{tabular}
\end{table}

To understand how developers perceive GitHub sponsors \textbf{(RQ3)}, we conducted a survey to get developer feedback on GitHub Sponsors.
In August 2021, we sent invitations to 2,000 randomly selected sponsored developers, 2,000 non-sponsored developers, and 2,000 sponsors to participate in the survey, and received a total of 372 responses: 177 from sponsored developers, 129 from non-sponsored developers, and 66 from sponsors.
Table~\ref{tab:participant} presents an overview of the demographics of our survey participants. Responses were obtained from GitHub users with diverse years of experience. Most of the respondents were male, but responses from other gender minorities were also included. Most respondents' contribution to OSS is part time, but sponsored developers are more likely to be full-time developers than others.

\subsection{Qualitative Analysis}
\label{ssec:qualitative}

For the qualitative analysis of the survey responses, three of the authors collaboratively took an initial look at all answers and discussed which themes were present in the data and how these themes related to the research questions. One of the authors then formalised this discussion into coding schemata.
Three authors applied the coding schemata to a random subset of the data, and one author finished the annotation based on the encouraging kappa agreements. We allowed multiple codes per answer.
In all cases, the kappa agreement was satisfactory after the first pass, and we did not change nor add codes during the process. We attribute this stability to the fact that we had an initial discussion about all data, that most answers were relatively short, and that this particular team of authors have experience working together on qualitative data analysis from previous research projects.
We describe each coding schema in the following questions. 

\textbf{Why are you looking for sponsors?}
To answer \textbf{RQ3.1}, we analyzed the reasons why OSS developers ask for donations. The three raters independently labeled 29 answers from two groups (from sponsored developers and non-sponsored developers). Then, we calculated the kappa agreement of our coding schema from three raters. Cohen's kappa for this qualitative analysis is 0.7, which indicates `substantial' agreement~\cite{viera2005understanding}. The following list shows our coding schema:
\begin{itemize}
    \item \textbf{funding a particular feature/ product}: We used this code if the respondent mentioned something in particular (other than the OSS project itself) that they were planning to use the donations for, e.g., ``\textit{Fund hosting costs for hosted projects.}''
    \item \textbf{gauge interest/satisfaction}: Some respondents indicated using sponsorship as a way to receive feedback from the community, in particular to assess the community's interest in the project, e.g., ``\textit{To measure the degree of satisfaction with my OSS works.}''
    \item \textbf{motivation}: We used this code if respondents were aware that receiving motivations would increase their motivation, e.g., ``\textit{It highly motivates me if I get money as donations for my work.}''
    \item \textbf{recognition/ appreciation}: If respondents expressed that they wanted to give the community a way to express appreciation or recognition, we used this code, e.g., ``\textit{Why not, mostly. Mostly boils down to recognition for the work since the monetary amounts are negligible.}''
    \item \textbf{financial support in general terms}: In cases where respondents gave a somewhat generic answer related to making money, we used this code, e.g., ``\textit{to fund my open source work.}''
    \item \textbf{none}: For respondents who did not mention a reason in response to the question, we used the code `none'.
    \item \textbf{not answered}: If the question was not answered, we coded it as `not answered'.
\end{itemize}

\textbf{Are you doing anything special to attract sponsors?}
To get a better understanding of what developers do to attract sponsors, we analyzed the responses of the actions that developers mentioned in their survey responses. The three raters independently coded 30 answers from two groups (from sponsored developers and non-sponsored developers), achieving Cohen's kappa of 0.91 or `almost perfect' agreement~\cite{viera2005understanding}. The following list shows the coding schema:
\begin{itemize}
    \item \textbf{perks for sponsors}: For responses which mentioned a specific perk that was only available to sponsors of their work, we used this code, e.g., ``\textit{I offer short consulting sessions (just a 30-60 minute video call) for higher tier sponsors.}''
    \item \textbf{social media}: We used this code if respondents mentioned a specific social media site or social media in general, e.g., ``\textit{Twitter - tweet about my work.}''
    \item \textbf{written content on GitHub}: Responses which mentioned updating documentation to make their work more attractive to sponsors, we used this code, e.g., ``\textit{I've written out my goals more publicly.}''
    \item \textbf{none}: If respondents did not mention any activities, we coded the response as `none'.
    \item \textbf{not answered}: If the question was not answered, we coded it as `not answered'.
\end{itemize}

\textbf{How has having sponsors affected you and the projects you are working on?/ How does the lack of sponsoring affect you and the projects you are working on?}
To answer \textbf{RQ3.2}, we analyzed the responses to the above questions of the impacts of (not) getting sponsorship to developers and OSS projects. Three raters independently annotated 28 answers from two groups (from sponsored developers and non-sponsored developers). The kappa agreement is 0.79, interpreted as `Substantial'~\cite{viera2005understanding}. Our coding schema is as follows:
\begin{itemize}
    \item \textbf{better project quality}: We used this code for responses which mentioned a positive impact of sponsoring towards specific quality attributes, e.g., ``\textit{It’s made me feel a lot more motivated towards adding more open source code, and paying a higher level of attention to detail.}''
    \item \textbf{limited effort}: This code was used for responses which mentioned a negative impact of the lack of sponsorship on the effort they were willing or able to spend, e.g., ``\textit{the projects receive less attention due to the need to do paid work.}''
    \item \textbf{motivation}: Responses mentioning the impact of (lack of) sponsorship on motivation received this code, e.g., ``\textit{Some pressure to finish the project.}''
    \item \textbf{satisfaction}: We used a similar category for responses related to satisfaction, e.g., ``\textit{Just for my heart's pleasure.}''
    \item \textbf{need other channels}: Responses which indicated that a lack of sponsorship via GitHub Sponsors implied that they used other sponsorship platforms fall into this category, e.g., ``\textit{I have people making contribution via other channels.}''
    \item \textbf{need other income}: Similarly, responses mentioning the need to look for other income in the absence of GitHub sponsorship were given its own code, e.g., ``\textit{I have to work on closed source and less interesting projects.}''
    \item \textbf{other}: For any answer to the question that did not fit the above categories, we used the code `other'.
    \item \textbf{none}: If respondents did not mention that there was any impact, we coded the response as `none'.
    \item \textbf{not answered}: If the question was not answered, we coded it as `not answered'.
\end{itemize}

\textbf{Why are you sponsoring others?/ Why did you become a sponsor?}
To answer \textbf{RQ3.3}, we analyzed the responses of the reasons why developers sponsor other OSS developers. Three raters independently coded 30 answers from our survey, achieving kappa of 0.62 or `substantial' agreement~\cite{viera2005understanding}. The lower agreement can be explained by 21 combinations of multiple codes being discovered when coding reasons for sponsoring. Three raters achieved perfect agreement in 15/30 cases (50\%) and partial agreement in another 14/30 cases (47\%), and completely disagreed only in 1/30 case (3\%). Our coding schema is as follows:
\begin{itemize}
    \item \textbf{dependencies/ benefited otherwise}: We used this code for responses which explicitly mentioned dependencies or other benefits, e.g., ``\textit{My project uses their projects.}''
    \item \textbf{excellent work}: If respondents indicated that they sponsored because of excellent projects, we used this code, e.g., ``\textit{I want to contribute to great projects.}''
    \item \textbf{form of involvement}: Related to the previous point, if respondents indicated that they see sponsoring as a form of involvement, we used this code, e.g., ``\textit{I had no more time to help with coding, so I decided to sponsor.}''
    \item \textbf{fairness/ give back to community}: We used this code if the main reason was related to fairness rather towards the community, e.g., ``\textit{If someone provides OSS software which I daily use, I think it's fair and important to pay a contribution.}''
    \item \textbf{set example}: Going a step further, some respondents mentioned that they sponsor to set an example for others, e.g., ``\textit{I want to set an example for other developers that large numbers of even small supporters can make an impact on keeping the OS community sustainable.}''
    \item \textbf{sustain project/ community}: If the response focused on using sponsorship to sustain a project or the community as a while, we used this code, e.g., ``\textit{To help them sustain their OSS software.}''
    \item \textbf{recognition/ appreciation}: If the responses mentioned recognizing a project or member of the community, or to show appreciation, we used this code, e.g., ``\textit{To praise their work, and recognize their invested time.}''
    \item \textbf{motivation}: Some respondents explicitly mentioned targeting the motivation of developers with their donations, e.g., ``\textit{Because I understand how hard it is to stay motivated.}''
    \item \textbf{required to get access to features/ training}: If participants mentioned sponsoring in return for specific access, we used this code, e.g., ``\textit{It was required to access training videos.}''
    \item \textbf{generic help/ support}: Somewhat generic statements such as ``\textit{Help valuable users in community}'' received this code.
    \item \textbf{other}: If the response did not fit any of these categories, we coded it as `other'.
    \item \textbf{not answered}: If the question was not answered, we coded it as `not answered'.
\end{itemize}

\textbf{What are the effects of your sponsorship? Are the effects as you expected?}
Lastly, we analyzed the responses of the effects of sponsoring, to the above questions. Three raters independently coded 27 answers from our survey. The kappa agreement is 0.82 or `almost perfect'~\cite{viera2005understanding}. We coded the responses according to the following categories:
\begin{itemize}
    \item \textbf{collective}: We used this code for responses which explicitly mentioned that their contributions on their own would not make a difference while the collective might, e.g., ``\textit{I do not expect to affect anyone with 1 USD a month. The collective does, but I personally don't have a say necessarily, which is good.}''
    \item \textbf{fund particular activities}: If respondents expected particular activities to take place in return for their sponsoring, we used this code, e.g., ``\textit{Maybe better handling of my files Issues. Also to keep a project alive.}''
    \item \textbf{grateful/ motivated developers}: Responses related to the motivation of the developers receiving sponsorship received this code, e.g., ``\textit{I'm not actually sure how to measure it. I just hope that provides enough signal to them to feel motivated.}''
    \item \textbf{recognition as sponsor}: We used this code for responses mentioning recognition of sponsors, e.g., ``\textit{Yes, as part of some sponsorships, we get our org some badges in their repo, to showcase our support.}''
    \item \textbf{required to get access to features/ training}: If the intended effect of sponsorship was to get access to certain features, we used this code, e.g., ``\textit{Getting involved in email threads and invitations to beta testing.}''
    \item \textbf{satisfaction}: Responses from sponsors mentioning their own satisfaction received this code, e.g., ``\textit{I honestly don’t get much out of sponsoring except personal satisfaction.}''
    \item \textbf{sustain project}: If the main effect was sustaining a particular project, we used this code, e.g., ``\textit{That keep the product alive and maintained.}''
    \item \textbf{none}: If respondents did not mention any effect, we coded it as `none'.
    \item \textbf{not answered}: If the question was not answered, we coded it as `not answered'.
\end{itemize}

\section{Results}

We present the answers to our research questions in this section.

\subsection{\RqOne}

To understand participants in GitHub Sponsors, we investigated the characteristics of sponsored developers and sponsors.

\paragraph{\textbf{RQ1.1: What are the characteristics of sponsored developers?}}

We measured the number of sponsors donating to the 3,697 sponsored developers in our dataset and found that 80\% had eight or fewer sponsors, and 1,355 (36\%) had just one sponsor. 661 (17\%) had 11 or more sponsors, and the developer with the highest number of sponsors had 1,342 sponsors. The median was 2.

\begin{table}
\caption{Primary programming languages of developers}
\label{tab:lang}
\small
\begin{tabular}{lrrrrr}
\toprule
& \multicolumn{2}{c}{\textbf{sponsored}} & \multicolumn{2}{c}{\textbf{non-sponsored}} & \textbf{GitHut} \\
\midrule
JavaScript & 796 & (22\%) & 1,240 & (22\%) & (19\%) \\
Python & 450 & (12\%) & 764 & (13\%) & (16\%) \\
PHP & 352 & (10\%) & 542 & (10\%) & (6\%) \\
C\# & 257 & (7\%) & 334 & (6\%) & (4\%) \\
TypeScript & 204 & (6\%) & 317 & (6\%) & (7\%) \\
Go & 203 & (5\%) & 298 & (5\%) & (8\%) \\
Java & 189 & (5\%) & 373 & (7\%) & (12\%) \\
C++ & 174 & (5\%) & 294 & (5\%) & (7\%) \\
C & 148 & (4\%) & 180 & (3\%) & (3\%) \\
Ruby & 130 & (4\%) & 169 & (3\%) & (7\%) \\
other & 794 & (21\%) & 1,155 & (20\%) & (11\%) \\
\midrule
\textbf{sum} & \textbf{3,697} & \textbf{(100\%)} & \textbf{5,666} & \textbf{(100\%)} & \textbf{(100\%)} \\
\bottomrule
\end{tabular}
\end{table}

\textbf{Primary programming languages.}
The main programming languages of the sponsored and non-sponsored developers identified by the method described in Section~\ref{ssec:data} are presented in Table~\ref{tab:lang}.
It can be seen that many sponsored developers contribute to JavaScript, Python, and PHP repositories.
Table~\ref{tab:lang} also shows the percentage of pull request volume by language for Q4 2020 taken from GitHut 2.0.\footnote{\url{https://madnight.github.io/githut/\#/pull_requests/2020/4}}
GitHut 2.0 is a web service that processes GitHub data with Google's BigQuery to visualize GitHub language statistics.\footnote{\url{https://github.com/madnight/githut}}
In all three sets of data, the set of programming languages in the top 10 is the same. This means that developers seeking sponsorship on GitHub Sponsors are in an ecosystem of programming languages that are actively being used.
JavaScript and Python are at the top of the three lists in common.
The other rankings for sponsored and non-sponsored developers are almost the same, but there are some differences between these and the GitHut ranking. Java, which is ranked third in terms of the volume of pull requests, is ranked seventh in terms of the main language used by sponsored developers. PHP, which ranks eighth in terms of the volume of pull requests, is found in third place as the primary language of sponsored developers.
We can see that there is a difference between the activity level of each programming language, as measured by the amount of pull requests, and the ecosystem of programming languages that have a presence in GitHub Sponsors.

\begin{figure}
    \centering
    \begin{subfigure}{.49\columnwidth}
        \centering
		\includegraphics[width=.7\linewidth]{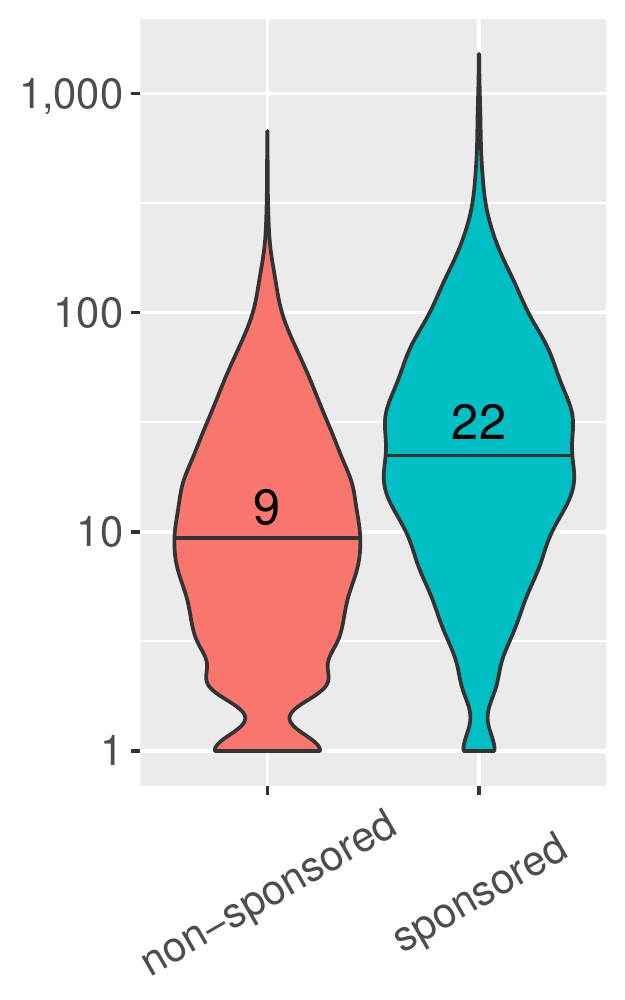}
		\caption{Number of repositories.}
	    \label{fig:repos}
    \end{subfigure}
    \begin{subfigure}{.49\columnwidth}
        \centering
        \includegraphics[width=.7\linewidth]{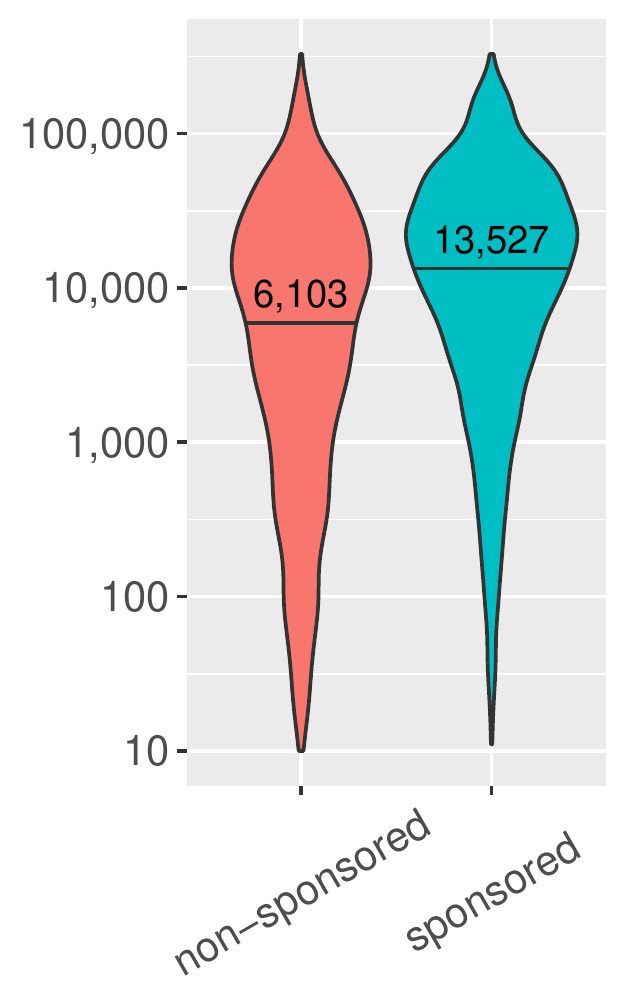}
		\caption{Highest number of stars in the repositories.}
	    \label{fig:star}
    \end{subfigure}
    \caption{Comparison of repositories contributed to by sponsored and non-sponsored developers.}
    \label{fig:RQ11}
\end{figure}

\textbf{Contributions to repositories.}
Figure~\ref{fig:repos} is a violin plot showing the number of repositories with 10 or more stars contributed to by sponsored and non-sponsored developers. The median values are 22 and 9, respectively.
The median number of repositories contributed to by the other 1,684,126 developers, excluding sponsored developers and non-sponsored developers (i.e., developers who had set up sponsorship, but not received donations), was 1. In this analysis of other developers, we excluded as many bot accounts as possible using keywords such as `[bot]', `-bot', `Bot', and manual checks.
Considering that most developers contribute to only one repository, we can see that both developer groups (sponsored and non-sponsored) are active, contributing to several popular projects, and especially sponsored developers contribute to many such projects.

Figure~\ref{fig:star} is a violin plot showing the highest number of stars in the repositories contributed to by sponsored and non-sponsored developers. The median values are 13,527 and 6,103, respectively.
We can see that sponsored developers contribute to repositories with a higher number of stars than non-sponsored developers.

\paragraph{\textbf{RQ1.2: What are the characteristics of sponsors?}}

\begin{figure}
    \centering
    \begin{subfigure}{.49\columnwidth}
        \centering
		\includegraphics[width=.5\linewidth]{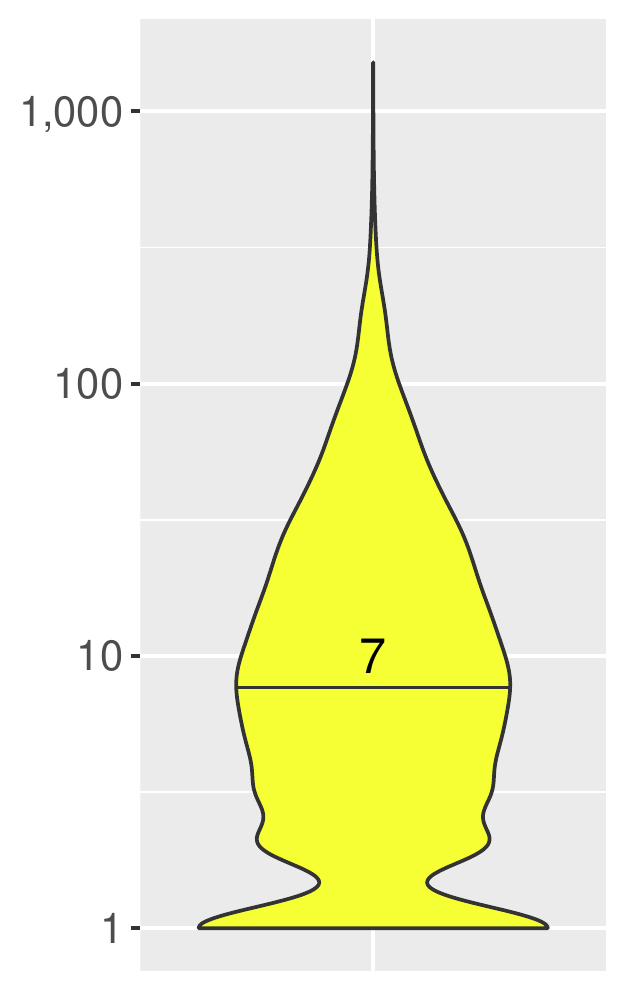}
		\caption{Number of repositories.}
	    \label{fig:repos_s}
    \end{subfigure}
    \begin{subfigure}{.49\columnwidth}
        \centering
        \includegraphics[width=.5\linewidth]{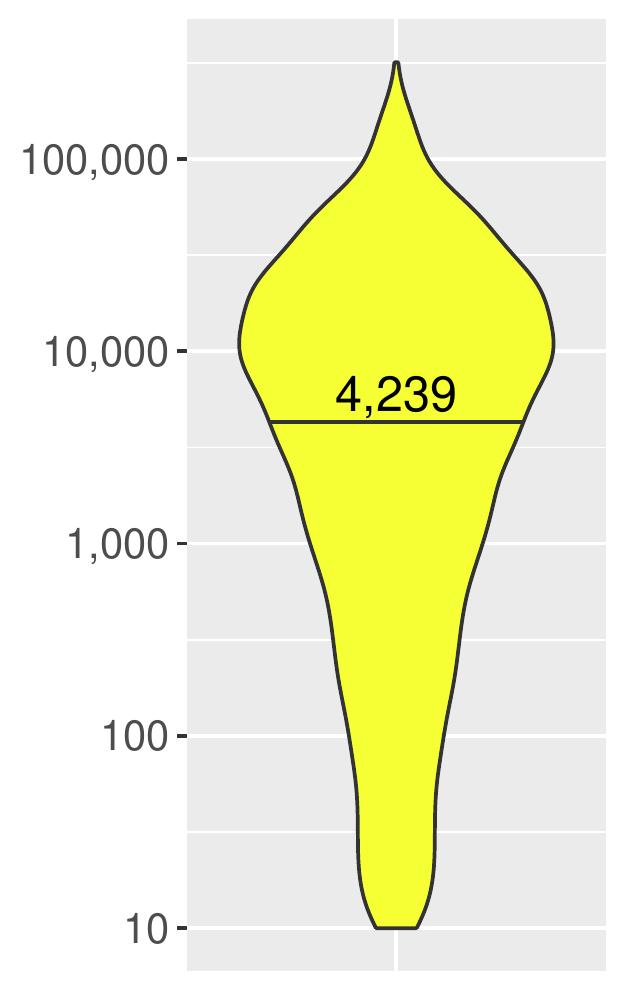}
		\caption{Highest number of stars in the repositories.}
	    \label{fig:star_s}
    \end{subfigure}
    \caption{Repositories contributed to by sponsors.}
    \label{fig:RQ12}
\end{figure}

Of the 17,458 sponsors we identified, 809 were sponsored developers themselves and 446 were non-sponsored developers. Of the remaining 16,203 sponsors who had not sought sponsorship for themselves, 6,020 (37\%) had not contributed to the repositories analyzed, while 10,183 (63\%) were developers who had contributed to repositories with 10 or more stars. From this, we found that two-thirds of the sponsors of GitHub Sponsors were developers, not just users.
Among GitHub Sponsors participants, of the 16,203 GitHub users who were only sponsoring, 80\% were only sponsoring one or two developers. Those who sponsored the most had sponsored 43 developers. The median 
was one.

\begin{figure}
  \centering
  \includegraphics[width=.7\linewidth]{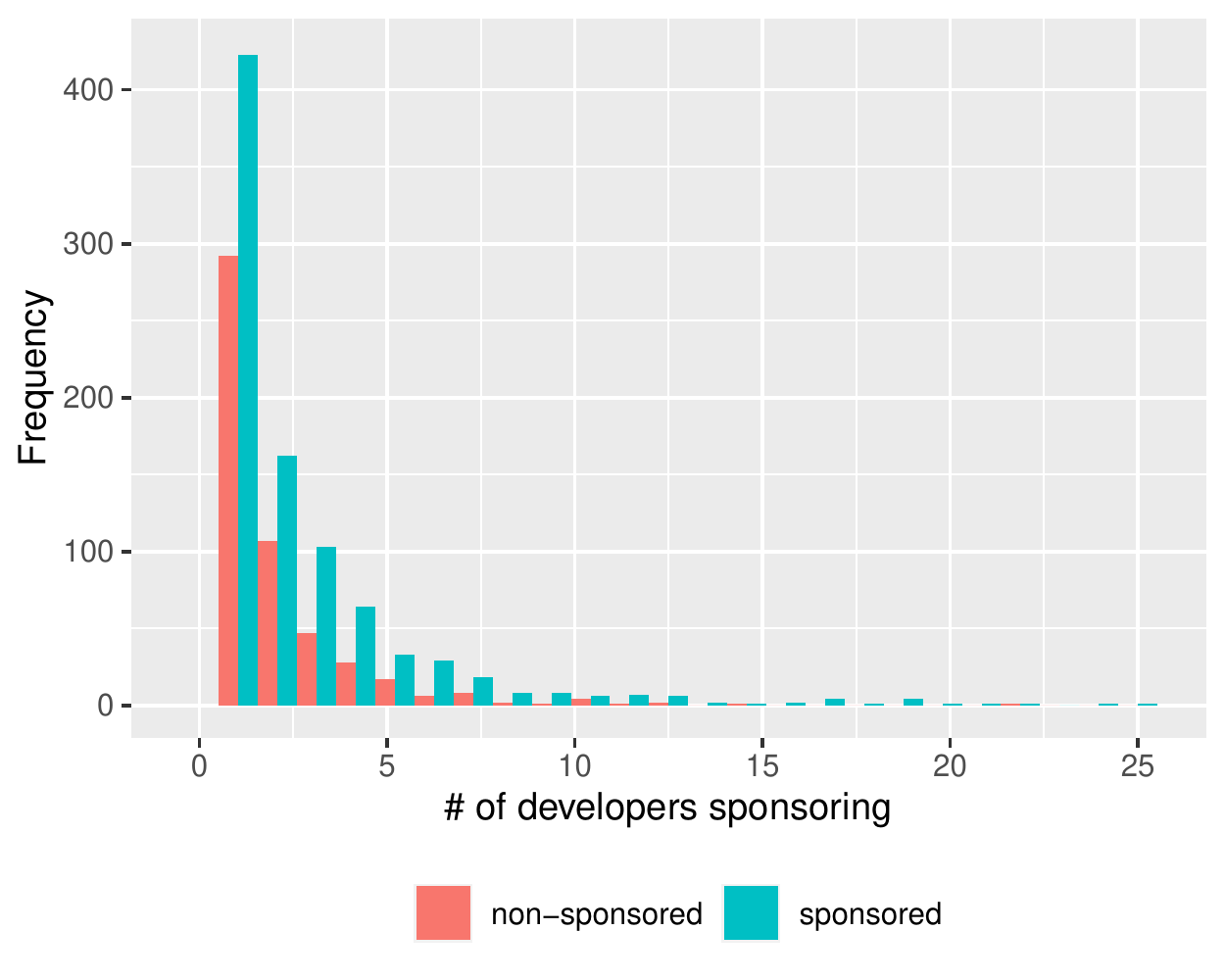}
  \caption{Distribution of the number of developers sponsored by other sponsored or non-sponsored developers.}
  \label{fig:sponsoring}
\end{figure}

\textbf{Contributions to repositories.}
Figure~\ref{fig:RQ12} presents a violin plot showing the number of repositories with 10 or more stars contributed to by 10,183 sponsors (\ref{fig:repos_s}) and a violin plot showing the highest number of stars in the repositories contributed to by the same sponsors (\ref{fig:star_s}).
From these figures, we can see that these sponsors are active developers, comparable to the non-sponsored developers seen in Figure~\ref{fig:RQ11}.

\textbf{Sponsors among sponsored and non-sponsored developers.}
Figure~\ref{fig:sponsoring} shows the distribution of the number of developers sponsored by other sponsored developers and non-sponsored developers. Most of them were sponsoring only a few developers, but the maximum values were 25 and 22, respectively.
We can see that the sponsored developer group has more developers sponsoring other developers than the non-sponsored developer group.

\begin{figure}
  \centering
  \includegraphics[width=\linewidth]{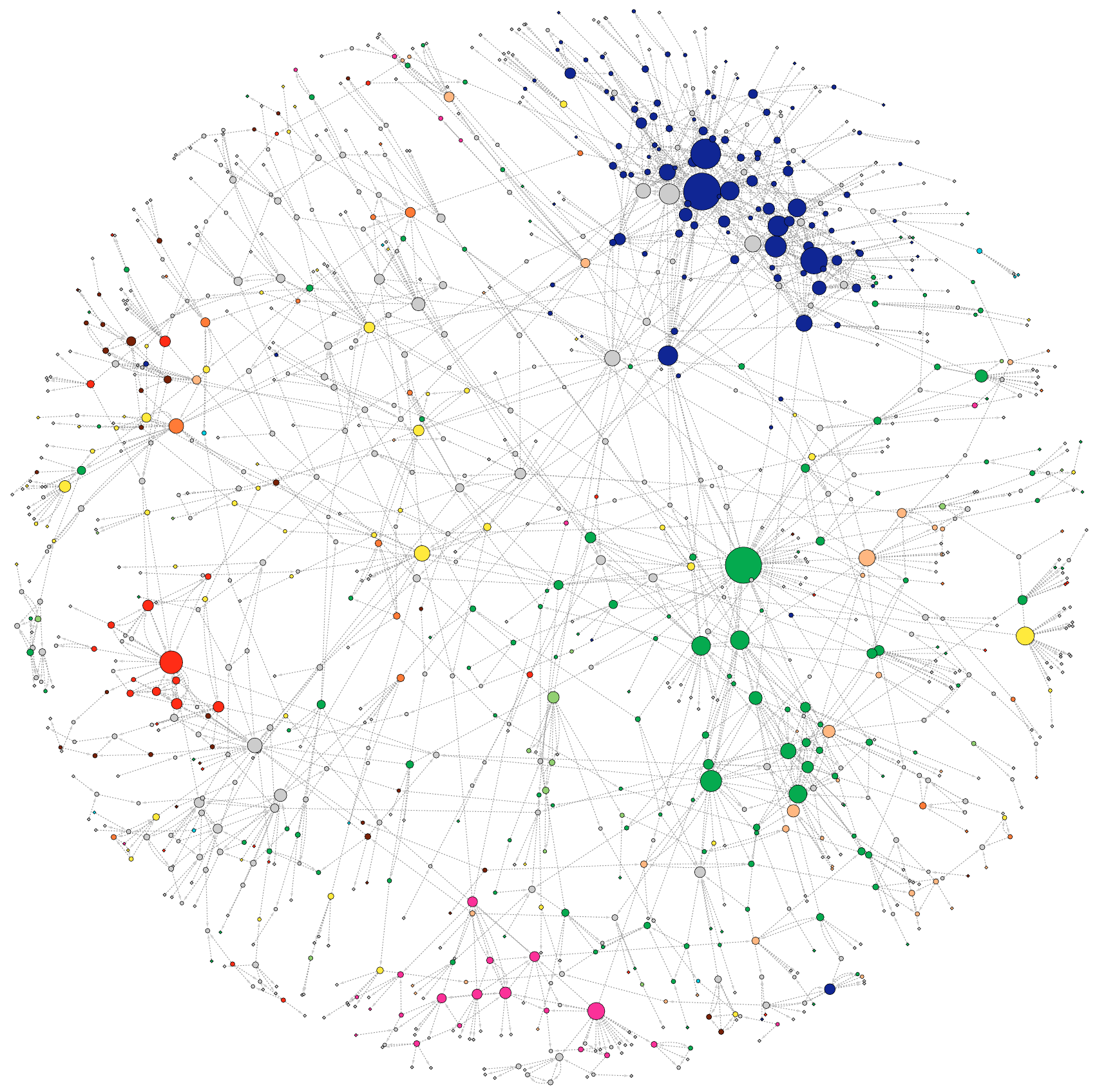}
  \caption{Largest network of sponsoring relationships among sponsored developers. Nodes represent developers and are color-coded by the top 10 major programming languages: \textcolor{green}{JavaScript}, \textcolor{yellow}{Python}, \textcolor{blue}{PHP}, \textcolor{magenta}{C\#}, \textcolor{apricot}{TypeScript}, \textcolor{brown}{Go}, \textcolor{yellowgreen}{Java}, \textcolor{orange}{C++}, \textcolor{skyblue}{C}, and \textcolor{red}{Ruby}.}
  \label{fig:sn}
\end{figure}

\textbf{Sponsoring network of sponsored developers.}
Of the 3,697 sponsored developers, 809 (22\%) were also sponsors, suggesting that developers are sponsoring each other. Therefore, we created networks with developers as nodes and sponsorship relationships as edges.
Figure~\ref{fig:sn} shows the largest of these networks.
The node size corresponds to the total number of inputs (being sponsored) and outputs (sponsoring) of the sponsorship. Developers whose primary language is one of the top 10 major programming languages listed in Table~\ref{tab:lang} are colored.
In this network, we can see that developers of the same major programming languages, such as JavaScript, PHP, C\#, and Ruby, have formed clusters.
In particular, we find that developers whose primary language is PHP form the closest and largest cluster, and they have a strong social network of donations through GitHub Sponsors.

\begin{tcolorbox}
\textbf{Summary}: Sponsored developers are more active than non-sponsored developers, with JavaScript, Python, and PHP being their top primary languages. About two-thirds of the sponsors are also active developers, and the sponsored developers form clusters that sponsor each other.
\end{tcolorbox}

\begin{figure*}
  \centering
  \includegraphics[width=.7\linewidth]{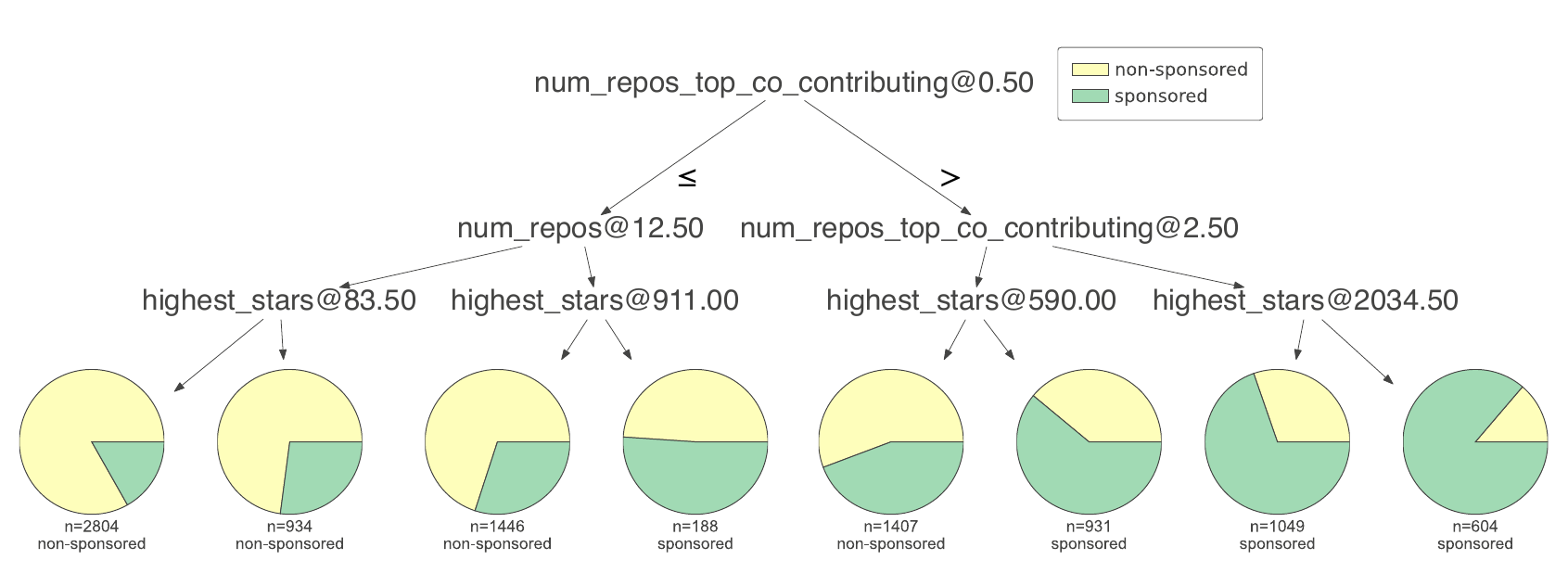}
  \caption{A decision tree for classifying sponsored and non-sponsored developers.}
  \label{fig:dt}
\end{figure*}

\subsection{\RqTwo}

To answer this research question, we employed a decision tree to examine the characteristics that are important in separating sponsored developers from non-sponsored developers. We also present developer feedback on important factors to consider when deciding who to sponsor.


\textbf{Decision tree.}
We use machine learning to find useful characteristics, or features, in the data.
Because the relationship between features and outcome could be nonlinear or the features may interact with each other, we conducted the analysis with decision trees rather than linear or logistic regression models.
The following features were measured for the outcome of having or not having sponsors (sponsored/non-sponsored). These features were designed based on the findings from the \textbf{RQ1} results. Note that repositories to be measured here had 10 or more stars.
\begin{itemize}
    \item \textbf{highest\_star}: Highest number of stars in the repositories that developers contributed to. As seen in Figure~\ref{fig:star}, sponsored developers are contributing to repositories with a high number of stars.
    \item \textbf{num\_repos}: Number of repositories that developers contribute to. As seen in Figure~\ref{fig:repos}, sponsored developers are contributing to many repositories.
    \item \textbf{num\_own\_repos}: Number of repositories contributed to and owned by a developer. Owning and managing popular projects might be a factor in getting sponsorship.
    \item \textbf{num\_repos\_top\_contributing}: Number of repositories contributed to that are ranked top in terms of commits. Making top contributions to popular repositories might be a factor in obtaining sponsorship.
    \item \textbf{num\_repos\_top\_co\_contributing}: Number of repositories contributed to with the top number of commits and that have at least one sponsor in the same repository. This does not take into account who the sponsors in the same repository are sponsoring. As seen in Figure~\ref{fig:sn}, sponsored developers form clusters of sponsoring relationships. The presence of GitHub Sponsors participants as sponsors in the development community might be a factor in obtaining sponsorship.
    \item \textbf{num\_sponsoring}: Number of sponsored developers. As seen in Figure~\ref{fig:sponsoring}, sponsored developers are sponsoring more than non-sponsored developers.
\end{itemize}

Figure~\ref{fig:dt} shows the result of applying the decision tree using scikit-learn\footnote{\url{https://scikit-learn.org/}} to the data of 9,363 developers, including 3,697 sponsored and 5,666 non-sponsored developers.
The feature \textit{num\_repos- \_top\_contributing} is found to be the most important, that is, the existence of such repositories seems to be the most common characteristic of sponsored developers.
If that value was three or more, most developers had obtained sponsors.
If that value was one or two and the feature \textit{highest\_star} was less than 590, more than half of the developers have not attracted sponsors.
If that value was zero, most developers did not get sponsors. However, if the number of contributing repositories is more than 13 and the feature \textit{highest\_star} is more than 911, then more than half of those developers had obtained sponsorship.
We found that sponsored developers not only contribute to the repositories with the most stars and contribute to the most repositories, but they are also the top contributors to the repositories where sponsoring developers are present. In other words, whether or not the communities to which developers belong have a habit of donating to OSS developers appears to be important for whether or not the developers get sponsors.

\textbf{Developer feedback.}
In the survey described in Section~\ref{ssec:survey}, we asked how important the following criteria were when deciding who to sponsor. The criteria are based on our insights gained from answering RQ1. The respondents were asked to choose from the following options: very important, somewhat important, a little bit important, not important at all, or don't care.
\begin{enumerate}[(a)] 
    \item They are major contributors to OSS projects that are important to me.
    \item They inspire me.
    \item We are contributing to the same OSS projects.
    \item We know each other outside the OSS community.
\end{enumerate}

\begin{figure}
  \centering
  \includegraphics[width=\linewidth]{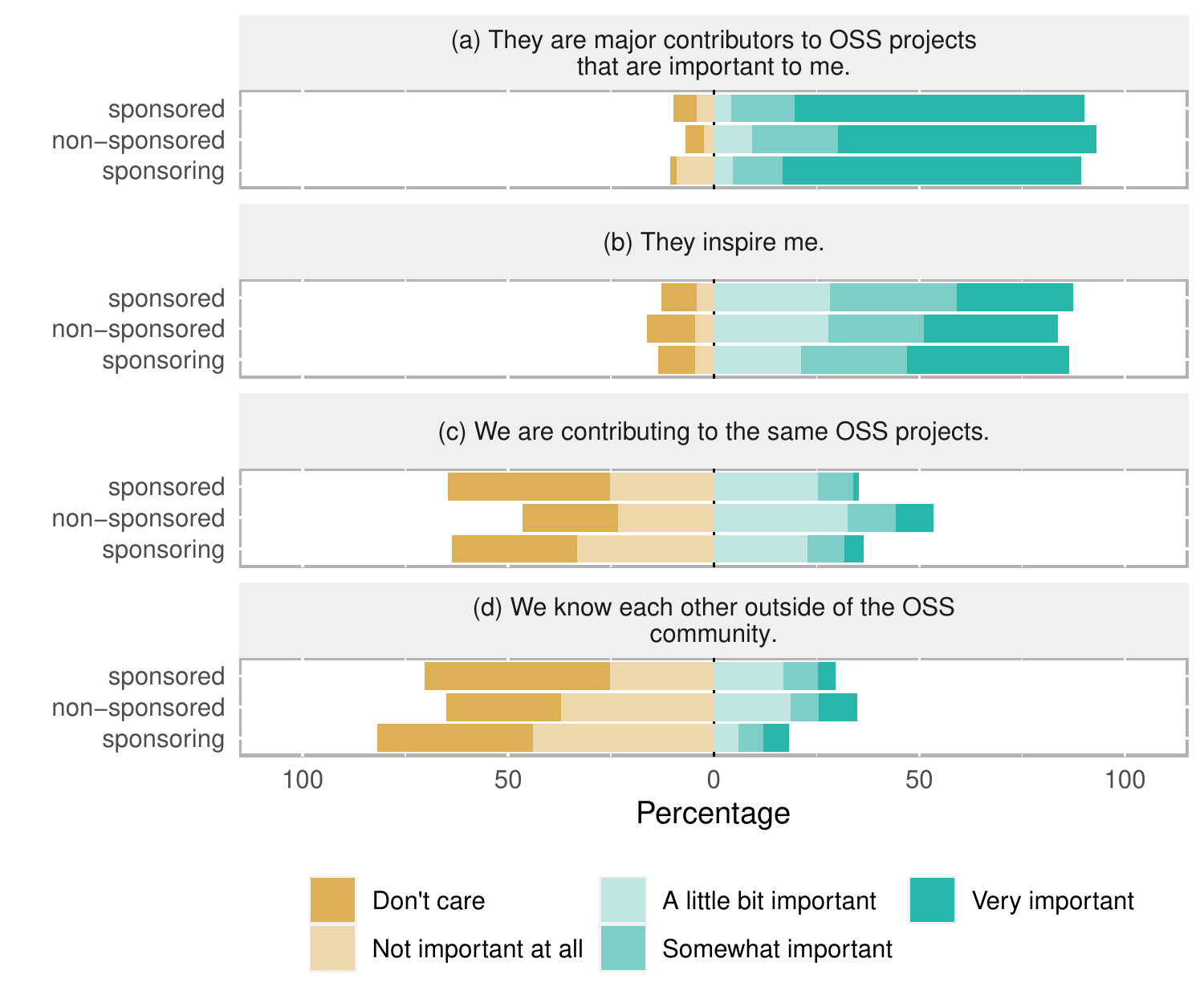}
  \caption{Distribution of the number of developers sponsored by other sponsored or non-sponsored developers.}
  \label{fig:survey}
\end{figure}

Figure~\ref{fig:survey} shows the distributions of the responses.
As expected, being a major contributor to an important OSS project (a) was a very important criterion.
In terms of the criterion of inspirational (b), the majority of responses were positive, including `very important', `somewhat important', and `a little bit important'. Because of the individual sponsorship, personal attractiveness seems to be important.
Frequent responses to the criterion of contributing to the same project (c) were `a little important', `not important at all', and `don't care'. It is not an important criterion for everyone, but it shows that some people do care about it. This can be considered as support for the importance of the feature \textit{num\_repos\_top\_co\_contributing} as revealed by the decision tree analysis.
Being an acquaintance (d) is mostly considered `not important at all' or `don't care'.

\begin{tcolorbox}
\textbf{Summary}: In addition to being a top contributor to popular repositories, we found that it is important to have developers sponsoring others in the development community. Our survey revealed that being a major contributor to an important OSS project and being inspirational were particularly important criteria in determining who to sponsor.
\end{tcolorbox}

\subsection{\RqThree}

To address this research question, we conducted qualitative analyses of the survey results, as described in Section~\ref{ssec:qualitative}.

\paragraph{\textbf{RQ3.1: Why are developers looking for sponsors?}}
Table~\ref{tab:q1} shows the result of our coding, separately for sponsored developers and non-sponsored developers.
From all 306 answers, the majority referred to financial support in general terms, accounting for 100 sponsored and 83 non-sponsored developers, respectively. Some sponsored developers started to accept donations to receive recognition/ appreciation of their works and effort, accounting for 37 answers. Among the reasons, to gauge interest/ satisfaction is a special reason related to the growth of OSS development. Developers consider GitHub Sponsors as a way to assess how the OSS community responds to their works. Some developers see GitHub Sponsors as a potential opportunity to become full-time OSS developers. For example, one respondent stated that ``\textit{I have enough money with my full time job, but I love contributing to OSS and giving to the community. Sponsors are mainly a way for me to see how much interest my projects get from the community. If it became self-sustaining some day, I could consider working full time on OSS.}''

\begin{table}
    \caption{Why are you looking for sponsors?}
    \label{tab:q1}
    \small
    \begin{tabular}{lrr}
    \toprule
    & \textbf{sponsored} & \textbf{non-sponsored} \\
    \midrule
    financial support in general terms  & 100 & 83\\
    recognition/ appreciation & 37 & 12\\
    funding a particular feature/product & 24 & 15\\
    motivation & 15 & 8\\
    none & 9 & 3 \\
    gauge interest/ satisfaction & 8 & 5\\
    not answered & 7 & 8\\
    \bottomrule
    \end{tabular}
\end{table}

\textbf{Are you doing anything special to attract sponsors?}
Table~\ref{tab:q2} shows the distribution of the codes across answers from sponsored developers and non-sponsored developers.
From all 306 answers, the majority of developers indicated no specific activities to attract sponsors, accounting for 106 and 103 respectively. Other actions included specific perks for sponsors and advertising on social media and/or the social coding platform itself.

\begin{table}
    \caption{Are you doing anything special to attract sponsors?}
    \label{tab:q2}
    \small
    \begin{tabular}{lrr}
    \toprule
    & \textbf{sponsored} & \textbf{non-sponsored} \\
    \midrule
    none & 106 & 103\\
    written content on GitHub & 36 & 10\\
    social media & 24 & 5\\
    perks for sponsors  & 11 & 4\\
    not answered & 7 & 8\\
    \bottomrule
    \end{tabular}
\end{table}

\begin{table}
    \caption{How has having sponsors affected you and the projects you are working on?/ How does the lack of sponsoring affect you and the projects you are working on?}
    \label{tab:q3}
    \small
    \begin{tabular}{lrr}
    \toprule
    & \textbf{sponsored} & \textbf{non-sponsored} \\
    \midrule
    none & 74 & 41\\
    motivation & 65 & 0\\
    better project quality  & 18 & 0\\
    satisfaction  & 14 & 0\\
    not answered & 14 & 10\\
    other  & 1 & 3\\
    limited effort & 0 & 61\\
    need other income  & 0 & 19\\
    need other channels & 0 & 1\\
    \bottomrule
    \end{tabular}
\end{table}

\paragraph{\textbf{RQ3.2: What is the impact of (not) getting sponsorship?}}
Table~\ref{tab:q3} shows the results of the qualitative analysis. 
%
We found that most of the sponsored developers (74) have not been affected by having sponsors. The second most frequently mentioned impact is motivation, accounting for 65 developers. For example, one respondent pointed out: ``\textit{It makes me feel bad when I ignore open pull requests or bug reports for too long, and causes me some anxiety as to whether it's really okay for me to ask for money, even though I make it very clear that sponsoring me won't make development faster or make me work on it full-time.}'' For non-sponsored developers, being able to dedicate only limited effort is the most common impact from not having sponsors, accounting for 61 developers.

\paragraph{\textbf{RQ3.3: Why are developers sponsoring?}}
Table~\ref{tab:q4} presents the results.
From the responses of 180 developers who indicated that they were sponsoring, we found that having dependencies on the code written by the receivers of the sponsorship and recognition/ appreciation are the most frequently mentioned reasons for all different roles of OSS developers.

\begin{table}
    \caption{Why are you sponsoring others?/ Why did you become a sponsor?}
    \label{tab:q4}
    \small
    \begin{tabular}{lrrr}
    \toprule
    & \textbf{sponsored} & \textbf{non} & \textbf{sponsoring} \\
    \midrule
    dependencies/ benefited otherwise  & 22 & 11 & 22\\
    recognition/ appreciation  & 22 & 13 & 17\\
    generic help/ support  & 11 & 7 & 14\\
    sustain project/ community & 8 & 4 & 3\\
    excellent work & 7 & 6 & 5\\
    fairness/ give back to community & 7 & 3 & 5\\
    motivation  & 5 & 4 & 4\\
    set example & 5 & 0 & 1\\
    not answered & 4 & 4 & 9 \\
    form of involvement & 2 & 0 & 4\\
    other & 0 & 1 & 1\\
    \multicolumn{3}{l}{required to get access to features/ training \hspace{5.1mm} 0 \hspace{6mm} 0} & 1\\
    \bottomrule
    \end{tabular}
\end{table}

\textbf{What are the effects of your sponsorship? Are the effects as you expected?}
Table~\ref{tab:q5} shows the distribution of the codes.
From all 180 answers, the majority of OSS developers have not observed any impact of their sponsorship. Other sponsors mentioned grateful/motivated developers, accounting for 13, 7, and 14 answers, respectively. One interesting theme mentioned in the responses is related to the collective of open source developers. For example, one respondent stated: ``\textit{No - It is difficult to know who to sponsor - and many people don't accept sponsorships. I wish folks who did not need money could gather and re-distribute money (i.e., flow through).}''

\begin{tcolorbox}
\textbf{Summary}: Our qualitative analysis revealed several benefits and challenges related to GitHub Sponsors, as perceived by its participants. While developers are hoping to receive financial support after signing up for GitHub Sponsors, the majority of developers does not do anything specific to attract sponsors. Sponsorship can afford funded developers more time to improve project quality, but those not receiving sponsorship tend to look elsewhere for money. The impact of sponsoring is not always clear to those making the donations, but many donate out of appreciation, not expecting anything specific in return.
\end{tcolorbox}

\begin{table}
    \caption{What are the effects of your sponsorship? Are the effects as you expected?}
    \label{tab:q5}
    \small
    \begin{tabular}{lrrr}
    \toprule
    & \textbf{sponsored} & \textbf{non} & \textbf{sponsoring} \\
    \midrule
    none & 31 & 20 & 27\\
    not answered & 13 & 8 & 16 \\
    grateful/motivated developers & 13 & 7 & 14\\
    satisfaction  & 8 & 4 & 1\\
    recognition as sponsor & 2 & 1 & 2\\
    fund particular activities & 2 & 0 & 2\\
    collective  & 2 & 0 & 0\\
    sustain project  & 1 & 4 & 3\\
    \multicolumn{2}{l}{required to get access to features/ training \hspace{5.1mm} 0} & 0 & 2 \\
    \bottomrule
    \end{tabular}
\end{table}

\section{Discussion}

In this section, we describe the threats to validity that affect our study, and we discuss implications and future work.

\subsection{Threats to validity}
\label{ssec:t2v}

\paragraph{Subjects are early adopters}
GitHub sponsorship started in 2019 and at this point, all participants are early adopters. There are not many developers who have obtained sponsors yet, which is a threat to the external validity of our empirical study. Therefore, it is important to note that the findings of this study are not generalizable to all open source developers, but rather results for early adopters.

\paragraph{Sponsored developers in less-starred repositories}
As explained in Section~\ref{ssec:data}, this study identifies users involved in GitHub Sponsors among developers who have contributed to repositories with 10 or more stars. Assuming that most sponsored developers contribute to popular repositories with a high number of stars, we targeted these repositories, but did not consider sponsored developers who only contribute to repositories with a low number of stars.
To examine the magnitude of the threat to external validity posed by this limitation, we analyzed a sample of repositories with fewer than 10 stars in the same way.
Due to the large number of less popular repositories on GitHub, it is not possible to retrieve all repositories with 9 or fewer stars from the API.
We used the list of repositories from RepoReapers, which examined 1,857,423 GitHub repositories for curating engineered software projects~\cite{10.1007/s10664-017-9512-6}, and investigated the star counts of all repositories in July 2021. We identified 1,338,533 repositories that had a star count of 9 or less.
From the developers who contributed to repositories with 9 or fewer stars, we excluded those who also contributed to repositories with 10 or more stars, indicated in Section~\ref{ssec:data}.
We obtained 450,884 contributors who contributed only to repositories with 9 or fewer stars.
Among these contributors, we identified seven sponsored developers.
We manually investigated the profiles of these seven people, and found that four had changed their GitHub account names, and that their accounts before the change were among the contributors to repositories with 10 or more stars.
Since there were only three sponsored developers out of the 450,884 developers who only contributed to repositories with 9 or fewer stars in the approximately 1.3 million repositories surveyed, we can conclude that the threat related to targeting contributors to repositories with 10 or more stars is small.

\paragraph{Up to 500 contributors from a repository}
When retrieving the contributors from a repository, only the top 500 developers with the most commits could be retrieved due to API limitations. However, a large repository can have more than 500 contributors.
Due to this limitation, developers who are not in the top 500 committers for at least one repository with 10 or more stars are not included in our analysis.
To examine the effect of this limitation, we measured the highest commit rank in repositories with 10 or more stars contributed to by each developer. The median and maximum values for sponsored developers were 1 and 351, respectively, while the median and maximum values for non-sponsored developers were 1 and 435, respectively.
Most sponsored and non-sponsored developers in our analysis contribute to at least one repository with 10 or more stars where they are the main committer, so the threat to external validity from the top 500 limit seems small.


\paragraph{Response biases}
Recall bias can occur in survey research. This means that respondents answer only what they remember and not necessarily what was most important to them in the past. Therefore, we tried to collect spontaneous responses and asked respondents to answer most of the questions in an open-ended format.

\subsection{Implications and Future Work}

GitHub Sponsors advertises their service as enabling ``a new way to contribute to open source.'' While donating to open source projects is not necessarily new, our study confirms that some sponsors do indeed see sponsorship as an alternative to contributing code, e.g., ``\textit{I had no more time to help with coding, so I decided to sponsor.}'' Most of the sponsors are active developers themselves, i.e., would likely have the technical skills to contribute in other ways as well. However, if they do not have sufficient time, sponsorship can be an effective way for them to show their support and appreciation.

A unique feature of GitHub Sponsors is its focus on enabling sponsorship for individual developers instead of entire projects. A side effect of this focus is that the social network of individual developers becomes an important determiner for whether they will be able to attract funding: our study identified several clusters in the network of sponsoring relationships among sponsored developers.
While we only investigated sponsorship by individual developers, organizational sponsorship is an important aspect that should be analyzed in future studies.

Developers who had signed up for receiving sponsorship seemed unsure what they could do to attract sponsors. It might be worth for GitHub to simplify the current sponsorship tier system which seems too complex for developers who are only looking for small donations to stay motivated. Expectations around what one gets in return for sponsorship are not always clear. From the perspective of sponsors, we identified a wide range of reasons for giving, from sustaining the entire OSS community to expecting particular features to be implemented. These expectations are not always perfectly matched with the reasons that developers enabled GitHub Sponsors in the first place.

A recent trend related to expectations and sponsorship is `sponsorware',\footnote{\url{https://github.com/sponsorware/docs}} a ``a release strategy for open-source software that enables developers to be compensated for their open-source work with fewer downsides than traditional open-source funding models.'' In those cases, expectations are made very explicit and software is not released until sponsorship has been obtained. This strategy stands in contrast to sponsors giving to show appreciation or to do something good for the OSS community in general. Future work will have to investigate whether sponsorware has the potential to disrupt the current OSS model.

\section{Related Work}
There are many factors that affect the sustainability of OSS projects. Researchers have argued that retaining newcomers is a way to sustain OSS projects~\cite{jensen2011joining, steinmacher2013newcomers}, and Steinmacher et al.~\cite{steinmacher2013newcomers} found that 20\% of newcomers became long-term contributors. Hata et al.~\cite{hata2015characteristics} suggest ways to incentivize developers to code and finally make projects sustainable (they assumed projects that can attract and retain coding contributors), which include that OSS projects should prepare documentation and make innovations to reduce code writing cost, and employ developers. Previous research~\cite{stewart2006impacts} showed that organizational sponsorship affects development activity over time.
According to GitHub's representative survey~\cite{github2017survey}, OSS development highly relies on voluntary contributions (23\% of respondents indicated they contribute to open source as part of their professional work). But recently, researchers found that the number of employees paid to work (they assumed work during weekdays from 9am---5pm as paid work) on OSS projects is increasing~\cite{riehle2014paid}. Moreover, Atiq and Tripathi~\cite{atiq2016impact} explored how developers perceive asymmetry in compensation in OSS projects, and found that OSS projects that are distributed unequally may fail if they are mismanaged. These researches point out that the financial benefits are a factor to sustain OSS projects.

\subsection{Donation}
Donations become a common way for obtaining these financial benefits~\cite{eghbal2019handy}.
Nakasai et al.~\cite{nakasai2017analysis,8501934} analyzed donations in the Eclipse projects, and found that benefits to donors and presenting badges could motivate donations. Krishnamurthy and Tripathi~\cite{RePEc:eee:respol:v:38:y:2009:i:2:p:404-414} found that commitment to an OSS platform contributes to the decision to donate. Moreover, donations decrease response time to bug reports of donors, and new releases are triggers of donations. Besides that, Overney et al.~\cite{10.1145/3377811.3380410} conducted a mixed-methods empirical study on the prevalence of donations, characteristics of projects requesting and receiving donations, and reasons for asking for donations in OSS projects. They found only 0.2\% of \texttt{npm} packages and 0.04\% of GitHub repositories ask for donations and the commonly used donation platforms are \texttt{PayPal} and \texttt{Patreon}. The projects asking for and receiving donations tend to be more active, more mature, and more popular. The authors also qualitatively analyzed donation information (e.g., \texttt{README.md} files and donation profile pages) and identified four reasons for asking for donations in OSS projects: Engineering (48\%), Community (18\%), Project expenses (13\%), and Personal (9\%). In addition, Yukizawa et al.~\cite{yukizawa2019please} promoted new strategies that applied social proof and legitimization of paltry contributions tactics to help OSS projects attract donations.

\subsection{Bounty}
Financial benefits are an important extrinsic motivation to sustain OSS projects~\cite{atiq2016impact}. Bounties are another way of obtaining these financial benefits~\cite{eghbal2019handy}. Unlike donations, bounties are used to motivate specific developers (i.e., bounty hunters) for tasks which are not of their primary interest by providing monetary incentives~\cite{kri2006bou,8170107}. 
Several studies investigated the effectiveness of bounty programs on disclosing vulnerabilities~\cite{finifter2013empirical,maillart2017given,zhao2017devising}.
Similar to Kanda et al.~\cite{7884685}, Zhou et al.~\cite{zhou2021studying} conducted an exploratory analysis on bounty backers and bounty hunters in the bounty issues addressing process of OSS projects from the \texttt{Bountysource} platform. They found that backers prefer to support implementing new features rather than fixing bugs, and bounty issues that are likely to be addressed by hunters tend to be in more popular and mature OSS projects. Moreover, the timing of proposing bounties is the most important factor in this process~\cite{zhou2020studying}. Zhou et al.~\cite{zhou2020bounties} also found that questions with bounty are likely to attract more traffic on Stack Overflow. Despite the advantages of bounties, they could also lead to disadvantages: winner-take-all property, disturbing the software development process, short-termism, alienating volunteer developers, bypassing existing leaders in the development community, attracting developers who lack project context, or redefining the priorities of the project in favor of the bounty~\cite{kri2006bou}.

\subsection{Motivation}
The motivation of OSS developers has been studied for decades. Lakhani and Wolf~\cite{lakhani2003hackers} grouped this motivation into three groups: 1) enjoyment-based intrinsic motivations, 2) obligation/community-based intrinsic motivations, and 3) extrinsic motivations. They found that being paid and
feeling creative on OSS projects does not have a significant negative impact on project effort, contrary to findings on the negative impact of extrinsic rewards on intrinsic motivations~\cite{deci1999meta}. Besides that, Roberts et al.~\cite{roberts2006understanding} argued that OSS developers being paid to participate leads to above-average contribution levels, while intrinsic motivations do not significantly impact average contribution levels. Hars and Ou et al.~\cite{927045} classified motivations of OSS developers into internal and external factors. They showed that external factors (e.g., building human capital, self-marketing, peer recognition, and personal needs) have greater weight.
Von Krogh et al.~\cite{von2012carrots} reviewed the literature on the motivation of OSS developers until 2009. They identified ten motivation categories grouped into intrinsic motivation, internalized extrinsic motivation, or extrinsic motivation. Gerosa et al.~\cite{gerosa2021shifting} investigated shifts in the motivation of OSS developers via a survey, and found that developers contribute to OSS projects for extrinsic factors, but continue contributing because of intrinsic factors. While financial benefits are common for experienced contributors, payment and money are still not common motivators. Moreover, the study of Krishnamurthy et al.~\cite{krishnamurthy2014acceptance} found that intrinsic and extrinsic motivations positively affect OSS developers' intention to accept monetary rewards.

\section{Conclusion}
To understand the potential of GitHub Sponsors in helping the open source community attract and retain developers, we conducted a mixed-methods study of 1,695,015 contributors from 1,168,856 repositories, focusing on sponsored developers, non-sponsored developers, and sponsors. We identified differences between those successful in attracting sponsorship and those still waiting for their first sponsor. The fact that developers are often clustered in terms of their sponsorship relationships implies that being connected to sponsors and other sponsored developers can have an impact on whether the search for sponsors is successful. While the impact of sponsorship is not always clear, and some sponsors were left disappointed that their donations did not have the intended effect, most sponsors contribute as a token of appreciation and recognition and/or for the good of the open source community.


\section{Data Availability}

Our online appendix contains the lists of studied repositories on GitHub, the dataset for the network diagram for answering RQ1, the features of sponsored and non-sponsored developers for RQ2, the features of sponsors for RQ2, and survey material and coding of responses for RQ3.
The appendix is available at at \url{https://github.com/NAIST-SE/GHSponsors} and \url{https://doi.org/10.5281/zenodo.6017901}.

\section{Acknowledgments}
This work has been supported by JSPS KAKENHI Grant Numbers JP20H00587 and JP20H05706.
\bibliographystyle{ACM-Reference-Format}
\bibliography{base}


\end{document}